\documentclass[12pt,a4paper]{article}
\usepackage[dvipdfmx]{graphicx}
\usepackage{amsmath,amssymb}
\usepackage{authblk}
\usepackage{comment}
\usepackage{here}
\usepackage{docmute}
\usepackage{simplewick}
\usepackage{cite}
\usepackage{layout}
\usepackage[margin=2.4cm]{geometry}
\usepackage{setspace} 
\usepackage{color}

\usepackage{ulem}
\usepackage{cancel}

\newcommand{\blue}[1]{\textcolor{blue}{#1}}

\newcommand{\bluewipe}[1]{\textcolor{blue}{\sout{#1}}}

\title{\bf The HAL QCD potential in $I=1$ $\pi \pi$ system with the $\rho$ meson bound state}
\author[1,2]{Yutaro Akahoshi}
\author[1,2]{Sinya Aoki}
\author[3,2]{Tatsumi Aoyama}
\author[2,4]{Takumi Doi}
\author[2]{Takaya Miyamoto}
\author[1,2]{Kenji Sasaki}
\affil[1]{\small Center for Gravitational Physics, Yukawa Institute for Theoretical Physics\\Kyoto University, Kyoto 606-8502, Japan}
\affil[2]{\small RIKEN Nishina Center (RNC), Saitama 351-0198, Japan}
\affil[3]{\small Institute of Particle and Nuclear Studies, High Energy Accelerator Research Organization(KEK)\\Tsukuba, Ibaraki 305-0801, Japan}
\affil[4]{\small RIKEN Interdisciplinary Theroretical and Mathematical Sciences Program (iTHEMS), Saitama 351-0198, Japan }

\begin{document}
\maketitle
\hspace{-0.6cm}\hrulefill
\begin{abstract}
In this paper, we investigate the HAL QCD potential in the $I=1$ $\pi \pi$ scattering using the hybrid method for all-to-all propagators, in which
a propagator is approximated by low-eigenmodes and the remaining high-eigenmode part
is stochastically estimated.
To verify the applicability of the hybrid method to systems containing quark creation$/$annihilation contributions such as the $\rho$ meson,
we calculate the {$I=1$ $\pi\pi$} potential with the 2+1 flavor gauge configurations on $16^3 \times 32$ lattice at the lattice spacing $a \approx 0.12$ fm and $(m_{\pi},m_{\rho}) \approx (870, 1230)$ MeV,
in which the $\rho$ meson appears as a deeply-bound state.
While we find that {the naive} stochastic evaluations
for quark creation$/$annihilation contributions lead to extremely large
statistical fluctuations,
additional noise reduction methods enable us to
obtain a sufficiently precise potential, which shows a strong attractive force.
We also confirm that the binding energy and $k^3 \cot \delta$ obtained from our potential are roughly consistent with an existing $\rho$ meson bound state, within a large systematic error associated with our calculation, whose possible origin is also discussed.
\end{abstract}
\hrulefill

\section{Introduction}
One of the most challenging issues in particle and nuclear physics is to understand hadronic resonances in terms of the fundamental theory of quarks and gluons, Quantum Chromodynamics(QCD).
To achieve this goal, two methods to study hadron-hadron interactions non-perturbatively in lattice QCD have been employed so far:
the L\"uscher's finite volume method~\cite{Luscher:1990ux,Rummukainen:1995vs,Hansen:2012tf} and the HAL QCD method~\cite{Ishii:2006ec,Aoki:2009ji,Aoki:2011ep,HALQCD:2012aa}.
{The L\"uscher's finite volume method enables us to calculate scattering phase shifts directly from finite-volume energy spectra.
Pole structures of bound states and resonances can be addressed
by the analytic continuation of the S-matrix into the complex energy region,
which however would require some ansatz for the structure of the S-matrix,
in particular for coupled channel systems.
Until now,
several mesonic resonances, such as the $\rho$ meson, have been studied in lattice QCD by this method~\cite{Briceno:2017max,Alexandrou:2017mpi,Werner:2019hxc}.}

In the HAL QCD method, on the other hand, an energy-independent but non-local potentials of hadron interactions are constructed from the Nambu-Bethe-Salpeter(NBS) wave function calculated in lattice QCD, from which physical observables are extracted afterward.
  This method has a unique advantage for the understanding of hadronic resonances from the first-principle.
  In this method, once the potential is obtained, one can directly address the pole structure
  of the S-matrix without any additional model-dependent ansatz.
  The extension to coupled channel systems,
  which are often essential to understand resonances,
  can be achieved in a straightforward manner~\cite{Aoki:2011gt}.
  Another strengh of this method is that the signal of the potential can be extracted
  not only from the ground state but also from excited states,
  which is crucial for reliable calculations for baryon-baryon systems~\cite{HALQCD:2012aa,Iritani:2018vfn}.
  Various interesting results have been reported in this method, for example,
  the identification of the $Z_c(3900)$ as the threshold cusp effect~\cite{Ikeda:2016zwx,Ikeda:2017mee}
  and predictions on the existence of $\Omega \Omega$ and $N \Omega$ di-baryons
  at the physical point~\cite{Gongyo:2017fjb,Iritani:2018sra}.

At present, however,
studies of resonances with the HAL QCD method are restricted due to the difficulty to treat all-to-all propagators within reasonable numerical costs and sufficient precisions.
In our previous attempts~\cite{Kawai:2017goq,Kawai:2018hem}, we utilized the LapH method~\cite{Peardon:2009gh} to treat all-to-all propagators, and it was revealed that the LapH smearing on the sink operator enhances the non-locality of the potential, so that the leading order approximation in the derivative expansion for the potential become insufficient.
To establish a more suitable way for all-to-all propagators, we have recently applied the hybrid method~\cite{Foley:2005ac}, which treats all-to-all propagators by the low-eigenmode approximation plus the stochastic estimation for the remaining high modes, to the HAL QCD method~\cite{Akahoshi:2019klc}.
In contrast to the LapH method, the hybrid method can keep the locality of quark operators since it contains full information of eigenmodes of the Dirac operator.
In Ref.~\cite{Akahoshi:2019klc}, we have studied the $I=2$ $\pi \pi$ S-wave scattering with the hybrid method, and we confirmed that the combination of the HAL QCD method and the hybrid method gave us
reliable results with better convergence of the derivative expansion,
as long as appropriate choices of parameters for the hybrid method have been made.

In this paper, we apply the hybrid method to $I=1$ $\pi \pi$ system
and study the $\rho$ meson.
Since all-to-all propagators are mandatory to calculate quark creation$/$annihilation contributions,
this is a best benchmark system to verify the applicability of the hybrid method.
We calculate the potential on the gauge configurations at $(m_{\pi},m_{\rho}) \approx (870,1230)$ MeV, in which the $\rho$ meson is not a resonance but a deeply-bound state.
It is revealed that stochastic estimations in the hybrid method for quark creation$/$annihilation contributions extremely enhance statistical fluctuations of the HAL QCD potential, and therefore we have to take some additional noise reductions to obtain a sufficiently precise potential.
As a consistency check, we calculate {the} binding energy and $k^3 \cot \delta$ {from} the resultant potential, and confirm that a deeply-bound $\rho$ state is reproduced within a somewhat large systematic error.

\vspace{5mm}
This paper is organized as follows.
In Sec. 2, we briefly explain the HAL QCD method and the hybrid method.
Simulation details in this study are given in Sec. 3.
Our main result, the potential of the $I=1$ $\pi \pi$ system, is presented in Sec. 4.
We also discuss physical observables computed by the potential and the origin of their systematic uncertainty here.
Our conclusion and outlook are given in Sec. 5.

\section{Method}
\subsection{HAL QCD method}
\indent
The fundamental quantity in the HAL QCD method is the Nambu--Bethe--Salpeter (NBS) wave function, which is defined for the $I=1$ two-pion system as
\begin{equation}
  \psi_{W}({\bf r},\Delta t) = \langle 0| (\pi \pi)_{I=1,I_z=0}({\bf r},0,\Delta t) |\pi \pi;I=1,I_z=0,{\bf k} \rangle,
\end{equation}
where $|\pi \pi;I=1,I_z=0,{\bf k} \rangle$ is an asymptotic state for the elastic $I=1$ $\pi \pi$ system in the center-of-mass frame with a relative momentum ${\bf k}$, the total energy $W = 2 \sqrt{m_{\pi}^2 + k^2}$ and $ k = \vert {\bf k} \vert$.
The operator $(\pi \pi)_{I=1,I_z=0}({\bf r},t,\Delta t)$ is a local two-pion operator projected to the $I=1, I_z=0$ channel, explicitly given by
\begin{equation}
  (\pi \pi)_{I=1,I_z=0}({\bf r},t,\Delta t) = \frac{1}{\sqrt{2}} \sum_{\bf x} \{ \pi^{+}({\bf r+x},t+\Delta t) \pi^{-}({\bf x},t) - \pi^{-}({\bf r+x},t+\Delta t) \pi^{+}({\bf x},t) \},
\end{equation}
where $\pi^{+}({\bf x},t)$($\pi^{-}({\bf x},t)$) is the positively (negatively) charged pion operator defined as
$\pi^{+}({\bf x},t) = \bar d({\bf x},t) \gamma_5 u({\bf x},t)$ ($\pi^{-}({\bf x},t) = \bar u({\bf x},t) \gamma_5 d({\bf x},t)$) with up and down quark fields $u({\bf x},t)$ and $d({\bf x},t)$.

The above definition of the NBS wave function is more general than the equal time ($\Delta t =0)$ NBS wave function, conventionally employed in the HAL QCD method,
where two sink hadron operators are put on the same time slice.
  In general, the HAL QCD potential depends on the choice of hadron operators in the definition of the NBS wave function, and we call it ``scheme''-dependence of the potential\cite{Kawai:2017goq,Aoki:2012tk}.
    The potential derived from the NBS wave function with $\Delta t \not=0$ belongs to
    the same scheme as the conventional equal time scheme
    if we take $\Delta t \rightarrow 0$ in the continuum limit,
    while it belongs to a different scheme if we keep physical $\Delta t$ finite in the continuum limit.
    Since calculations in this study are performed only at one lattice spacing, 
    we consider the results from $\Delta t= 0$ and $\Delta t \not=0$ as
    those in two different schemes between which the discretization artifact appears differently.

    While the potentials are scheme-dependent,
    physical quantities such as phase shifts and binding energies,
    of course, do not depend on the scheme (up to the discretization errors).
    On can even take advantage of this arbitrariness
    by choosing a better scheme so that statistical/systematic errors are minimized.
As discussed later, the main reason why we introduce the
scheme with non-zero $\Delta t$
is to reduce statistical fluctuations of the potential for the $I=1$ $\pi\pi$ system,
which are caused by stochastic estimations for all-to-all quark propagators in the hybrid method.

As discussed in Ref.~\cite{Aoki:2009ji,Aoki:2013cra} for  the case of the $\Delta t=0$ scheme,
we can show the radial part of the  $l$-th partial component in the NBS wave function with the non-zero $\Delta t$ scheme
behaves at large $r = \vert{\bf r}\vert$ as
\begin{equation}
  \psi^{l}_{W}(r,\Delta t) \approx A_{l}(\Delta t,{\bf k}) e^{i\delta_l} \frac{\sin(kr-l \pi/2+\delta_l(k))}{kr},
\end{equation}
where $A_{l}(\Delta t,{\bf k})$ is an overall  factor and $\delta_l(k)$ is the scattering phase shift,
which is equal to  the phase of  the S-matrix implied by its unitarity.
By using this behavior, we can construct an energy-independent but non-local potential through the Schr\"odinger-type equation as
\begin{equation}
  \frac{1}{2 \mu}(\nabla^2 + k^2) \psi_{W}({\bf r},\Delta t) = \int d^3 {\bf r'}\, U_{\Delta t}({\bf r},{\bf r'})\psi_{W}({\bf r'},\Delta t),
\end{equation}
where $\mu = m_{\pi}/2$ is a reduced mass of two-pions, and a subscript $\Delta t$ of $U$ represents the scheme for the potential. In practice, the non-locality of the potential is treated by the derivative expansion as
\begin{equation}
  U_{\Delta t} ({\bf r},{\bf r'}) = (V_{\Delta t}^{\rm LO} (r) + V_{\Delta t}^{\rm NLO}(r) \nabla^2 + {\mathcal O}(\nabla^4)) \delta({\bf r-r'}).
\end{equation}

The normalized $\pi\pi$ correlation function, numerically calculable in lattice QCD, is related to the NBS wave functions as
\begin{equation}
R({\bf r},t,\Delta t) \equiv \frac{F({\bf r},t,\Delta t)}{C(t)^2} \approx \sum_n {B_n} \psi_{W_n}({\bf r},\Delta t) e^{-(W_n - 2m_{\pi}) t} + ...
\end{equation}
{
where $W_n$ and $B_n$ are the energy and overlap factor of the $n$-th excited elastic state, and an ellipsis indicates inelastic contributions.
}
Here $C(t)$ and $F({\bf r},t,\Delta t)$ are $\pi$ and $\pi\pi$ correlation functions defined by
\begin{eqnarray}
  C(t) &=& \sum_{{\bf x,y},t_0}\langle \pi^{-}({\bf x},t+t_0) \pi^{+}({\bf y},t_0) \rangle \\
  F({\bf r},t,\Delta t) &=& \sum_{t_0} \langle (\pi \pi)_{I=1,I_z=0}({\bf r},t+t_0,\Delta t) {\mathcal J}^{T_1^-}_{I=1,I_z=0}(t_0) \rangle,
\end{eqnarray}
where ${\mathcal J}^{T_1^-}_{I=1,I_z=0}(t_0)$ is a source operator which creates $I=1$ and $I_z=0$ $\pi \pi$ scattering states in the $T_1^-$ representation.
Among several choices for the source operator, we take  a $\rho$-type source operator in our study, given by
\begin{equation} \label{eq:rho-src}
  {\mathcal J}^{T_1^-}_{\rho;I=1,I_z=0}(t_0) = \sum_{\bf x} \bar \rho^0_3 ({\bf x},t),
\end{equation}
where $\rho^0_i$ is the neutral $\rho$ meson operator, $\rho^0_i = \bar u \gamma_i u - \bar d \gamma_i d$.
{Since this source operator strongly overlaps {with} the $\rho$ meson state,
  {we expect that
    the truncation error of the derivative expansion in the effective leading-order analysis
    is suppressed around the mass of the $\rho$ meson.
  }
}

The normalized correlation function $R({\bf r},t,\Delta t)$ satisfies~\cite{HALQCD:2012aa}
\begin{equation} \label{eq:timedepHAL}
  \left[ \frac{\nabla^2}{m_{\pi}} -\frac{\partial}{\partial t} + \frac{1}{4m_{\pi}} \frac{\partial^2}{\partial t^2} \right] R({\bf r},t,\Delta t) = \int d^3{\bf r'} U_{\Delta t}({\bf r},{\bf r'}) R({\bf r'},t,\Delta t),
\end{equation}
at a sufficiently large $t$ where inelastic contributions in $R({\bf r},t,\Delta t)$ becomes negligible.
From eq.(\ref{eq:timedepHAL}), the effective leading-order(LO) potential is obtained as
\begin{equation}
  V_{\Delta t}^{\rm LO}(r) = \frac{\left[ \dfrac{\nabla^2}{m_{\pi}} -\dfrac{\partial}{\partial t} + \dfrac{1}{4m_{\pi}} \dfrac{\partial^2}{\partial t^2} \right] R({\bf r},t,\Delta t)}{R({\bf r},t,\Delta t)}.
\end{equation}
Using the rotational invariance of the potential,
we can rewrite the above definition to improve signals as~\cite{Murano:2013xxa}
\begin{equation}
  V_{\Delta t}^{\rm LO}(r) = \frac{ \sum_{g\in O_h} R^{\dag}(g{\bf r},t,\Delta t) \left[ \dfrac{\nabla^2}{m_{\pi}} -\dfrac{\partial}{\partial t} + \dfrac{1}{4m_{\pi}} \dfrac{\partial^2}{\partial t^2} \right] R(g{\bf r},t,\Delta t)}{\sum_{g\in O_h} R^{\dag}(g{\bf r},t,\Delta t) R(g{\bf r},t,\Delta t)},
\end{equation}
where the $O_h$ is the cubic rotation group.
{We also note that we employ the 4th order difference approximation for $\nabla^2$
to reduce discretization errors at short distances, since it turns out that  physical observables in the deeply-bound system are sensitive to the potential at short distances. }

\subsection{All-to-all propagator: the hybrid method}
In this subsection, we briefly explain the hybrid method, a technique
for the all-to-all propagator calculation  employed in this study.
Let us consider
the spectral decomposition of the quark propagator as
\begin{equation}
  D^{-1}(x,y) = \sum_{i=0}^{N-1} \frac{1}{\lambda_i} v^{(i)}(x) \otimes v^{\dag(i)}(y) \gamma_5,
\end{equation}
where $v^{(i)}(x)$ and $ \lambda_i$ are eigenvectors and eigenvalues of the Hermitian Dirac operator $H=\gamma_5 D$, respectively, with  $N$ being the total number of eigenmodes,
and  color and spinor indices are implicit for simplicity.
We here assume $|\lambda_i| \le |\lambda_j|$ for $i < j$.

\begin{figure}[tbp]
  \centering
  \includegraphics[width=60mm,clip]{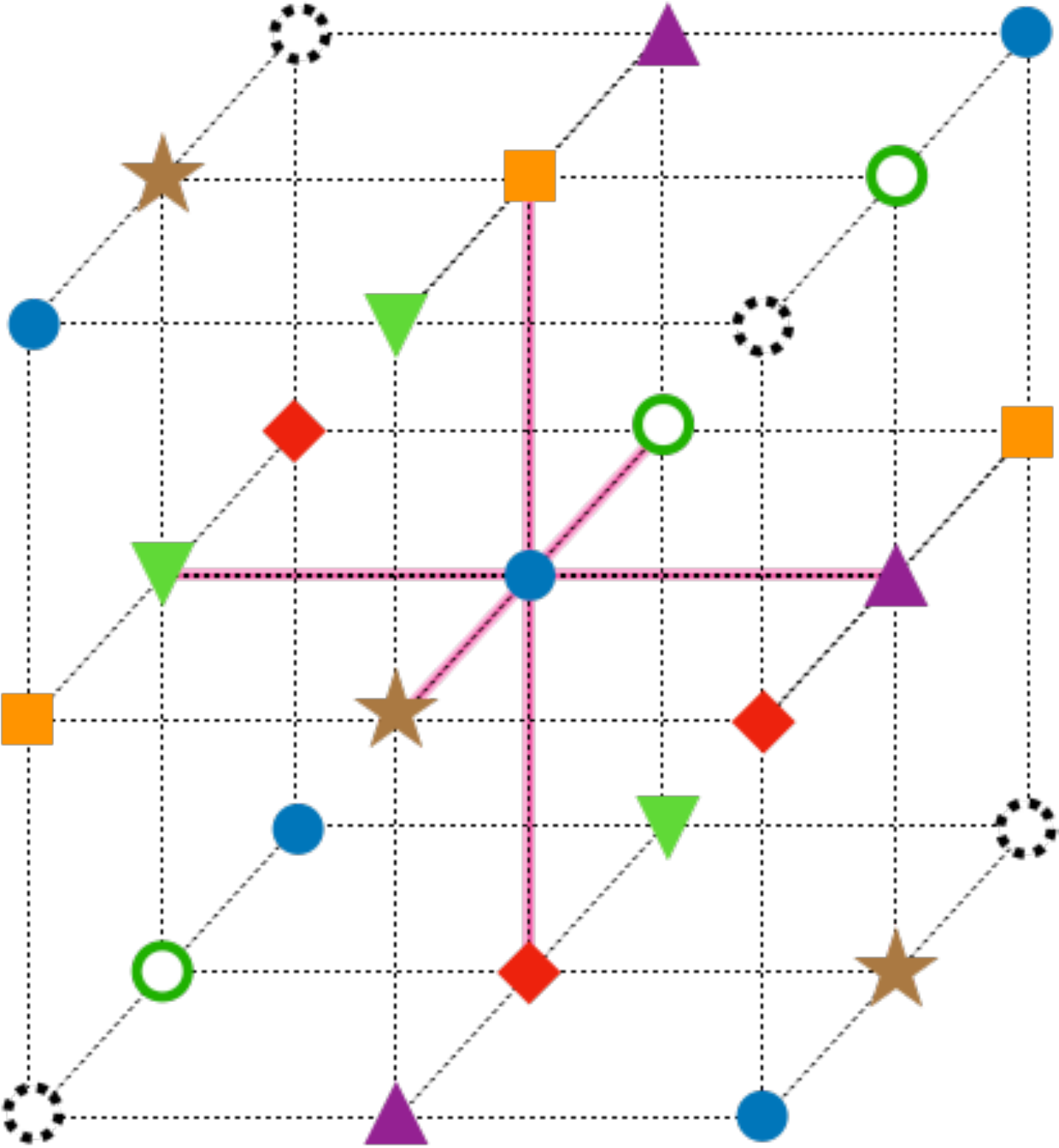}
  \caption{Schematic representation of $s8$ space dilution. different symbols indicate points {which belong} 
  to different diluted vectors. Pink bands connect points used in the 2nd order approximation of the Laplacian at the central point.}
  \label{fig:s8dilution}
\end{figure}

The low-eigenmode approximation for the propagator with the spectral decomposition is introduced as
\begin{equation}
  D_0^{-1}(x,y) = \sum_{i=0}^{N_{\rm eig}-1} \frac{1}{\lambda_i} v^{(i)}(x) \otimes v^{\dag(i)}(y) \gamma_5, \quad  N_{\rm eig}<N,
\end{equation}
while
the remaining high-eigenmode part is estimated by using the $Z_4$ noise vector $\eta_{[r]}^{(i)}$,  together with the variance reduction by dilution as
\begin{equation}
  D^{-1} - D_0^{-1} = H^{-1} {\mathcal P}_1 \gamma_5 \approx \frac{1}{N_{\rm r}} \sum_{r=0}^{N_{\rm r}-1} \sum_{i=0}^{N_{\rm dil}-1} \psi_{[r]}^{(i)}(x) \otimes \eta_{[r]}^{\dag(i)}(y) \gamma_5,
\end{equation}
where ${\mathcal P}_1 \equiv {\bf 1} - \sum_{i=0}^{N_{\rm eig}-1} v^{(i)} \otimes v^{\dag(i)}$ is a projection onto the remaining high-eigenmode part, $N_{\rm r}$ ($N_{\rm dil}$) is a number of noise vectors (dilutions),
and $\psi_{[r]}^{(i)}$ are solution vectors obtained by solving $ H\cdot \psi_{[r]}^{(i)}= {\mathcal P}_1 \eta_{[r]}^{(i)}$.
In this study, the temporal coordinate is diluted  with the $J$--interlace as
\begin{equation}
  \eta^{(i)}({\bf x}, t) \not= 0, \quad  \mbox{only if $t= i$ mod $J$}.
\end{equation}
For spatial coordinates, we introduce not only $s2$ (even-odd) and $s4$ dilutions used in the previous study~\cite{Akahoshi:2019klc}, but also a $s8$ dilution.
In the $s8$ dilution, one noise vector is split into 8 diluted vectors as
\begin{eqnarray}
  \begin{array}{ccc}
   \eta^{(0)} \neq 0 & \ {\rm if} \begin{cases}
     \mbox{$(n_x,n_y,n_z)$ = (odd,odd,odd) and $n_x+n_y+n_z = 1 \ {\rm mod}\ 4$ } \\
     \mbox{$(n_x,n_y,n_z)$ = (even,even,even) and $n_x+n_y+n_z = 2 \ {\rm mod}\ 4$ }
   \end{cases}\\\vspace{1mm}
   \eta^{(1)} \neq 0 & \ {\rm if} \begin{cases}
     \mbox{$(n_x,n_y,n_z)$ = (odd,odd,odd) and $n_x+n_y+n_z = 3 \ {\rm mod}\ 4$ } \\
     \mbox{$(n_x,n_y,n_z)$ = (even,even,even) and $n_x+n_y+n_z = 0 \ {\rm mod}\ 4$ }
   \end{cases}\\\vspace{1mm}
   \eta^{(2)} \neq 0 & \ {\rm if} \begin{cases}
     \mbox{$(n_x,n_y,n_z)$ = (odd,even,even) and $n_x+n_y+n_z = 1 \ {\rm mod}\ 4$ } \\
     \mbox{$(n_x,n_y,n_z)$ = (even,odd,odd) and $n_x+n_y+n_z = 2 \ {\rm mod}\ 4$ }
   \end{cases}\\\vspace{1mm}
   \eta^{(3)} \neq 0 & \ {\rm if} \begin{cases}
     \mbox{$(n_x,n_y,n_z)$ = (odd,even,even) and $n_x+n_y+n_z = 3 \ {\rm mod}\ 4$ } \\
     \mbox{$(n_x,n_y,n_z)$ = (even,odd,odd) and $n_x+n_y+n_z = 0 \ {\rm mod}\ 4$ }
   \end{cases}\\\vspace{1mm}
   \eta^{(4)} \neq 0 & \ {\rm if} \begin{cases}
     \mbox{$(n_x,n_y,n_z)$ = (even,odd,even) and $n_x+n_y+n_z = 1 \ {\rm mod}\ 4$ } \\
     \mbox{$(n_x,n_y,n_z)$ = (odd,even,odd) and $n_x+n_y+n_z = 2 \ {\rm mod}\ 4$ }
   \end{cases}\\\vspace{1mm}
   \eta^{(5)} \neq 0 & \ {\rm if} \begin{cases}
     \mbox{$(n_x,n_y,n_z)$ = (even,odd,even) and $n_x+n_y+n_z = 3 \ {\rm mod}\ 4$ } \\
     \mbox{$(n_x,n_y,n_z)$ = (odd,even,odd) and $n_x+n_y+n_z = 0 \ {\rm mod}\ 4$ }
   \end{cases}\\\vspace{1mm}
   \eta^{(6)} \neq 0 & \ {\rm if} \begin{cases}
     \mbox{$(n_x,n_y,n_z)$ = (even,even,odd) and $n_x+n_y+n_z = 1 \ {\rm mod}\ 4$ } \\
     \mbox{$(n_x,n_y,n_z)$ = (odd,odd,even) and $n_x+n_y+n_z = 2 \ {\rm mod}\ 4$ }
   \end{cases}\\\vspace{1mm}
   \eta^{(7)} \neq 0 & \ {\rm if} \begin{cases}
     \mbox{$(n_x,n_y,n_z)$ = (even,even,odd) and $n_x+n_y+n_z = 3 \ {\rm mod}\ 4$ } \\
     \mbox{$(n_x,n_y,n_z)$ = (odd,odd,even) and $n_x+n_y+n_z = 0 \ {\rm mod}\ 4$ }
   \end{cases}
  \end{array}.
\end{eqnarray}
See Fig.~\ref{fig:s8dilution} for
a schematic figure of the $s8$ dilution. Color and spinor indices are fully diluted in this study.

Combining the low-eigenmode and the high-eigenmode parts, the all-to-all propagator is written as
\begin{equation}
  D^{-1} \approx \frac{1}{N_{\rm r}} \sum_{r=0}^{N_{\rm r}-1} \sum_{i=0}^{N_{\rm hl}-1} u_{[r]}^{(i)} \otimes w_{[r]}^{\dag (i)} \gamma_5,
\end{equation}
where the hybrid lists $u_{[r]}^{(i)},w_{[r]}^{(i)}$ are defined by
\begin{eqnarray}
  w_{[r]}^{(i)} &=& \{ \frac{v^{(0)}}{\lambda_0}, \cdots , \frac{v^{(N_{\rm eig}-1)}}{\lambda_{N_{\rm eig}-1}},\eta_{[r]}^{(0)}, \cdots ,\eta_{[r]}^{(N_{\rm dil}-1)} \} \\
  u_{[r]}^{(i)} &=& \{ v^{(0)}, \cdots , v^{(N_{\rm eig}-1)}, \psi_{[r]}^{(0)}, \cdots ,\psi_{[r]}^{(N_{\rm dil}-1)} \}
\end{eqnarray}
with $N_{\rm hl} = N_{\rm eig} + N_{\rm dil}$.

\subsection{
  Correlation function
  with the hybrid method}
The correlation function $F({\bf r},t,\Delta t)$ with the $\rho$-type source operator,
\begin{equation}
  F({\bf r},t,\Delta t) = \sum_{{\bf y},t_0} \langle (\pi \pi)_{I=1,I_z=0}({\bf r},t+t_0,\Delta t) \bar \rho^0_3({\bf y},t_0) \rangle,
\end{equation}
is expressed in terms of the hybrid method  (up to an overall sign) as
\begin{equation}
  \begin{split}
    \sum_{i,j,k} \sum_{{\bf x},t_0} &O^{(i,j)}_{[r,s]}({\bf r+x},t+t_0+\Delta t) O^{(j,k)}_{3[s,p]}(t_0) O^{(k,i)}_{[p,r]}({\bf x},t+t_0) \\
    &- O^{(i,j)}_{[r,s]}({\bf x},t+t_0) O^{(j,k)}_{3[s,p]}(t_0) O^{(k,i)}_{[p,r]}({\bf r+x},t+t_0+\Delta t),
  \end{split}
  \label{eq:contraction}
\end{equation}
where
\begin{equation}
  O^{(i,j)}_{3[r,s]}(t) \equiv \sum_{{\bf x}} w^{\dag (i)}_{[r]} ({\bf x},t) \gamma_5 \gamma_3 u^{(j)}_{[s]}({\bf x},t) , \quad
 O^{(i,j)}_{[r,s]}({\bf x},t) \equiv  w^{\dag (i)}_{[r]} ({\bf x},t) u^{(j)}_{[s]}({\bf x},t).
\end{equation}
Note that equal--time quark propagators would appear due to contractions in the sink operator if we took $\Delta t = 0$  in the calculation.

\section{Simulation details}
\label{sec:details}
In this study, we employ 2+1 flavor full QCD configurations generated by JLQCD and CP-PACS Collaborations~\cite{Ishikawa:2007nn} on a $16^3 \times 32$ lattice with the Iwasaki gauge action\cite{Iwasaki:1985we} at $\beta=1.83$ and a non-perturbatively improved Wilson-clover action\cite{Sheikholeslami:1985ij} at $c_{SW} = 1.7610$ and hopping parameters $(\kappa_{ud},\kappa_s) = (0.1376,0.1371)$.
These parameters correspond to
the lattice spacing $a = 0.1214$ fm, the pion mass $m_{\pi} \approx 870$ MeV, and the $\rho$ meson mass $m_{\rho} \approx 1230$ MeV. Note that the $\rho$ meson is not a resonance but a bound state
of two pions in this calculation. The periodic boundary condition is employed for all spacetime directions.

Tab.~\ref{tab:setups} shows details of our numerical setup,
{whereas parameters for the hybrid method are summarized in Tab.~\ref{tab:hybridsetups}.}
{
  In case 0, the source operator in eq.~(\ref{eq:rho-src}) is constructed
  from the point quark source.
  In case 1, on the other hand,
}
we employ the smeared quark source $q_s({\bf x},t) = \sum_{\bf y} f({\bf x-y})q({\bf y},t)$ with the Coulomb gauge fixing, 
{so that inelastic contributions are reduced at earlier imaginary times.}
The smearing function $f$ is given by~\cite{Iritani:2017rlk}
\begin{equation}
  f \left( {\bf x}  \right) = \begin{cases}
    a e^{-b |{\bf x }|} & ( \ 0 < |{\bf x }| < (L-1)/2 \ ) \\
    1 & ( \ |{\bf x }| = 0 \ ) \\
    0 & ( \ |{\bf x }| \geq (L-1)/2 \ )
  \end{cases}
\end{equation}
with $a=1.0,b=0.47$ in lattice unit.
%
%
As regards the setup for the random noise vectors, case 0 is calculated with
  three independent $Z_4$ noise vectors corresponding to $r,s,p$ in eq.~(\ref{eq:contraction}).
  In case 1, 
  we generate four different sets of three $Z_4$ noise vectors, and take an average over $4 \times 3! =24$ samples ($3! = 6$ samples for each set using the permutation of $r,s,p$) to reduce noise contamination.
Statistical errors are estimated by the jackknife method with bin--size 1 (6) in case 0 (case 1).

\begin{table}[tbp]
  \caption{{Numerical setup} for the calculations.}
  \vspace{2mm}
  \centering
  \begin{tabular}{c|cccc}
    & Source & Scheme & $N_{\rm conf}$ & Stat. error   \\ \hline \hline
    case 0 & point & equal-time ($\Delta t = 0$) & 20 & jackknife with binsize 1 \\
    case 1 & smear & different-time ($\Delta t = 1$) & 60 & jackknife with binsize 6
  \end{tabular}
  \label{tab:setups}
\end{table}
\begin{table}[tbp]
  \caption{Setups for the hybrid method in our calculation. $N_{\rm eig}$ is the number of low eigenmodes for the all-to-all propagator. Color and spinor dilutions are always used.}
  \vspace{2mm}
  \centering
    \begin{tabular}{c|ccc}
      & time dilution & space dilution & $N_{\rm eig}$  \\ \hline \hline
      case 0 & 16-interlace & $s2$ & 100  \\
      case 1 (src-to-sink) & 16-interlace & $s4$ & 100 \\
      case 1 (sink-to-sink) & 4-interlace & $s8 \times s2$ & 100
    \end{tabular}
    \label{tab:hybridsetups}
\end{table}

In case 0,
  we employ the equal-time ($\Delta t = 0$) scheme.
  As will be shown in Sec.~\ref{sec:simple}, however,
  the statistical errors of the potential are found to be
  too large to obtain physical results,
  probably due to the
  statistical fluctuations associated with
  the equal--time quark propagations in the sink operator.
We therefore employ $\Delta t = 1$ scheme in case 1
and avoid equal--time quark propagations.
In addition,
we make a spatial dilution finer in the sink--to--sink propagator to reduce noise contamination in the Laplacian part, whose increased numerical costs are partly compensated by
decreasing the temporal dilution from the 16--interlace to the 4--interlace.
Since we found
in the previous study~\cite{Akahoshi:2019klc}
that propagations along the temporal direction from $t_0$ to $t_0 + t$ with $t < J/2$ are not distorted much by the $J$--interlace dilution,
the 4--interlace temporal dilution reduces the computational cost for sink--to--sink propagations
without additional strong noise enhancements.

\begin{figure}[tbp]
  \hspace{-10mm}
  \begin{tabular}{cc}
  \begin{minipage}{0.5\hsize}
    \includegraphics[width=85mm,clip]{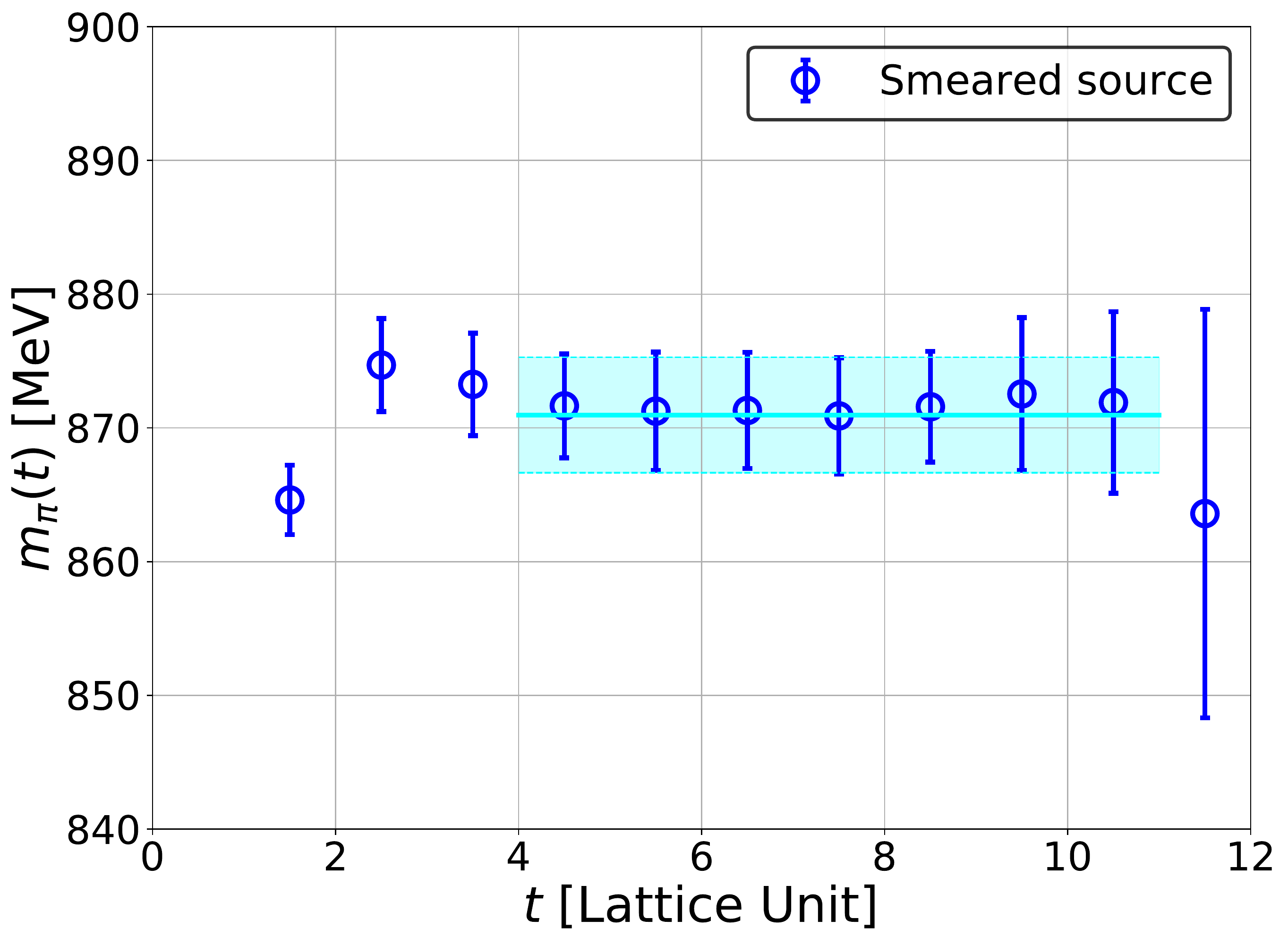}
  \end{minipage} &
  \begin{minipage}{0.5\hsize}
    \includegraphics[width=85mm,clip]{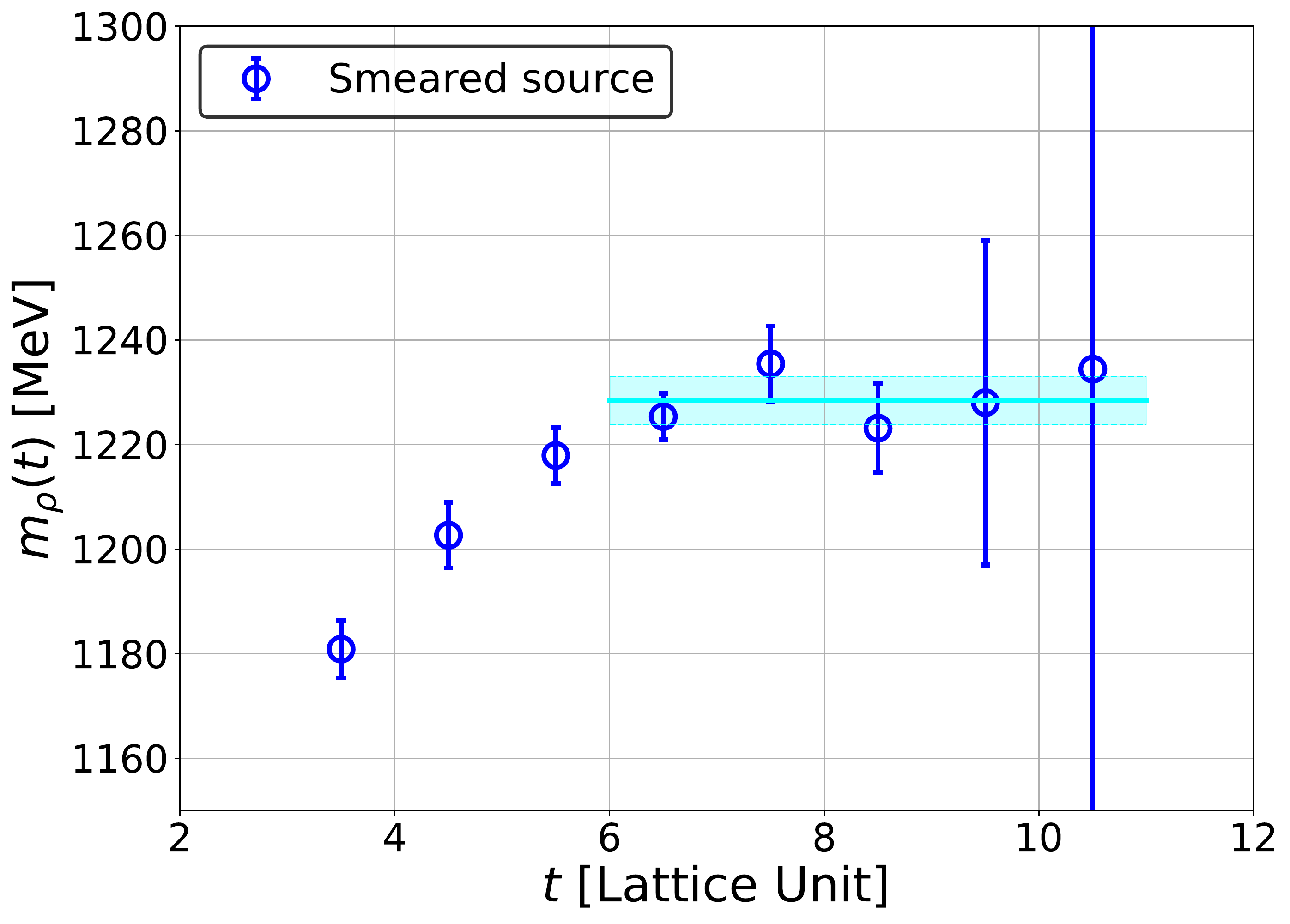}
  \end{minipage}
\end{tabular}
\caption{(Left) The effective mass of pion. (Right) The effective mass of $\rho$ meson. {Both effective masses are obtained in case 1.} Each cyan solid line with band indicates the central value and its statistical error obtained from the fit to the corresponding correlation function within this interval.}
\label{fig:effmasses}
\end{figure}

Fig.~\ref{fig:effmasses} {(Left) and (Right)} show the effective masses
  {(with a half-integer time convention~\cite{Akahoshi:2019klc}) obtained in case 1}
  {for pion $m_{\pi}(t)$ and $\rho$ meson $m_{\rho}(t)$,}
  which are calculated from $C(t)$ and $F(t) \equiv \sum_{\bf r} \bar Y_{l=1,m=0}(\Omega_{\bf r}) F({\bf r},t,\Delta t)$, respectively.
Note that we insert the spherical harmonics {for the P-wave} $\bar Y_{l=1,m=0}$ in the summation
to obtain $F(t)$,
{which is relevant to the $\rho$ meson.}
The fit to $C(t)$ at $t=4-11$ gives $m_{\pi} = 871(4)$ MeV, while the fit to $F(t)$ at $t=6-11$ gives $m_{\rho} = 1228(5)$ MeV.
The ratio of $m_{\pi}$ and $m_{\rho}$ becomes $m_{\pi}/m_{\rho} = 0.709(4)$, which is consistent with
{$m_{\pi}/m_{\rho} = 0.7076(18)$}
reported in the previous study~\cite{Ishikawa:2007nn}.
Fig.~\ref{fig:effmasses}  also shows that  the ground state saturations in $C(t)$ and
$F(t)$
are achieved at least $t=4$ and $t=6$, respectively, {in case 1}.
  In case 0, while the ground state saturation in $C(t)$ is achieved at later time than case 1
  (see Fig. 2 (Left) in \cite{Akahoshi:2019klc}),
  $t=6$ is found to be sufficient 
  since errors in the potential is dominated by the statistical fluctuations
  as will be shown in Sec.~\ref{sec:simple}.
In the following, we take results at $t=6$ as our central values and
use results at $t=5,7$ to estimate systematic errors associated with their time dependence.

\section{Results}
\subsection{
Potential in case 0
}
\label{sec:simple}
\begin{figure}[tbp]
  \hspace{-10mm}
  \begin{tabular}{cc}
  \begin{minipage}{0.5\hsize}
    \includegraphics[width=85mm,clip]{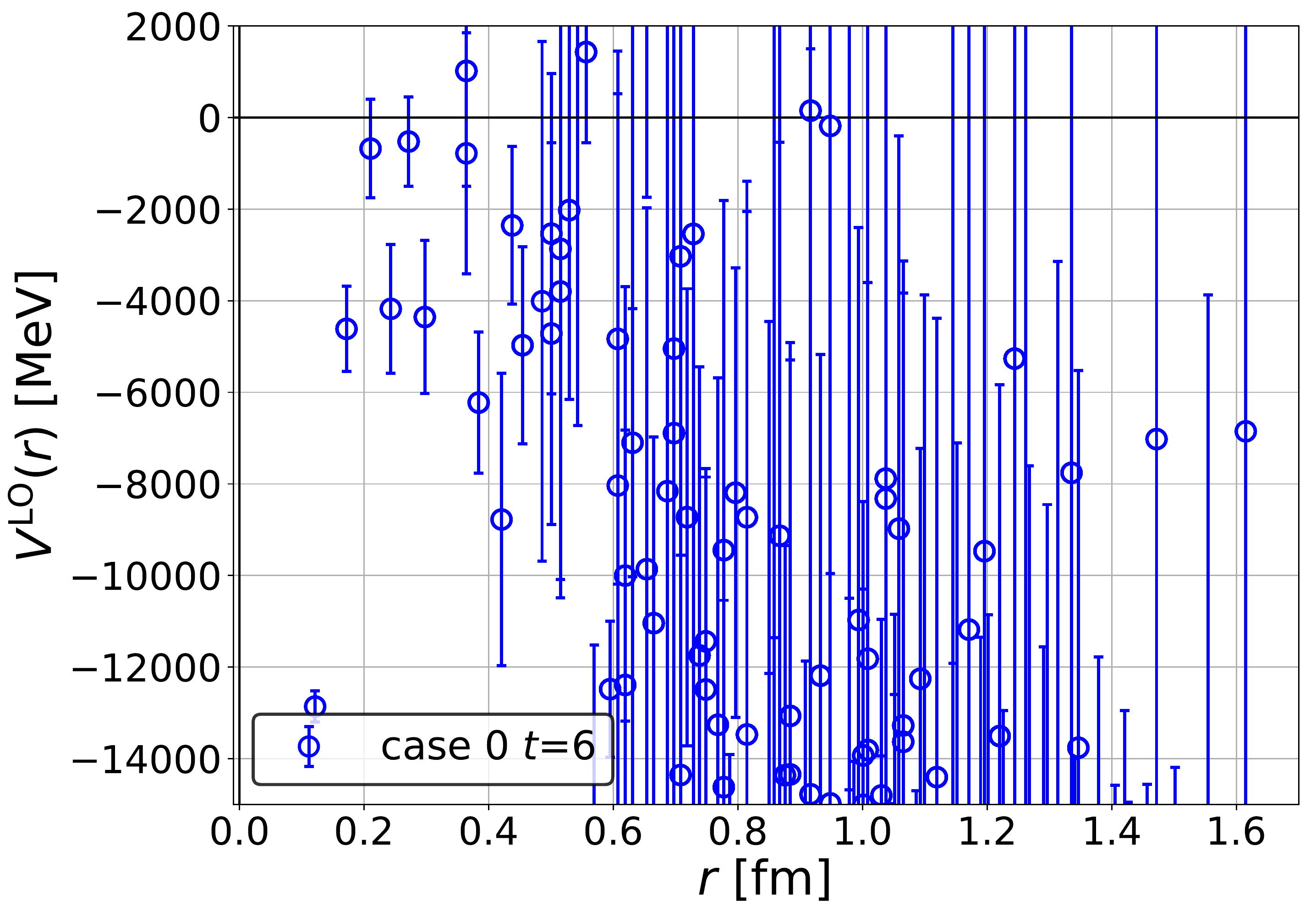}
  \end{minipage} &
  \begin{minipage}{0.5\hsize}
    \includegraphics[width=85mm,clip]{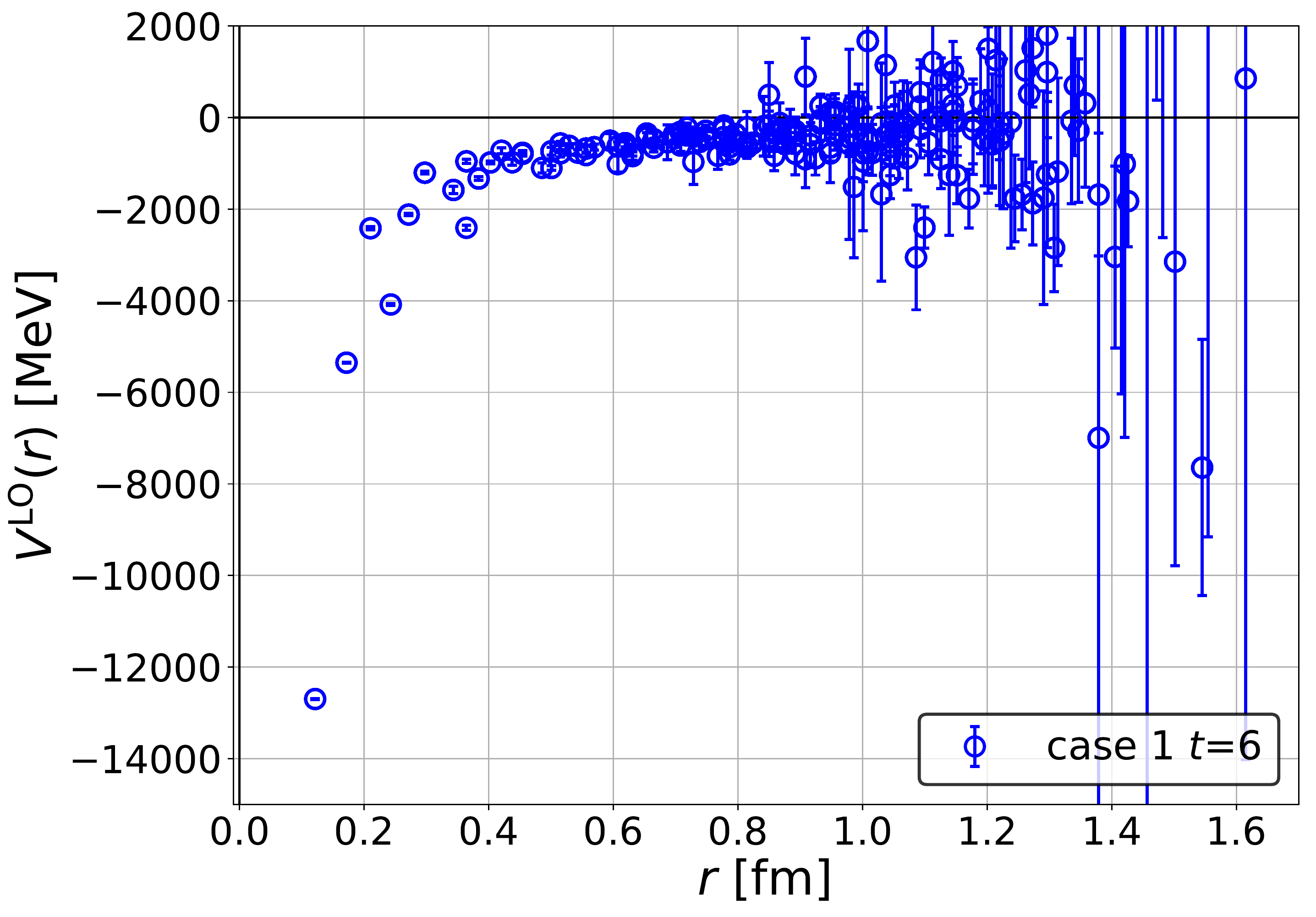}
  \end{minipage}
\end{tabular}
\caption{(Left) The potential  at $t=6$ in case 0 (the same setup as case 3 in Ref.~\cite{Akahoshi:2019klc}). (Right) The potential at $t=6$ in case 1.}
\label{fig:potential}
\end{figure}

We first consider the case 0 for the $I=1$ $\pi \pi$ potential,
whose setup for the hybrid method is the same as the case 3 for the $I=2$ $\pi \pi$ potential
in Ref.~\cite{Akahoshi:2019klc}.
In the previous study, we have found that the $I=2$ $\pi \pi$ potential is reasonably accurate at $t<8$.
Fig.~\ref{fig:potential}(Left) shows the potential obtained at $t=6$.
As can be seen, the potential has  extremely large statistical fluctuations in this setup.
Since equal--time quark propagations at the sink were absent for the $I=2$ $\pi \pi$ potential
in  the previous study,
we suspect that
extremely large statistical fluctuations for the $I=1$ $\pi \pi$ potential
are caused by noise contaminations from the hybrid method to evaluate
such equal--time propagations at the sink.

To suppress such noise contaminations, we additionally employ three noise reduction techniques, (1) the different--time scheme for the NBS wave function to avoid the equal--time propagation, (2) the finer space dilution in the quark annihilation part to reduce noise contamination in spatial indices, (3) the average over different noise vectors.
In the following, we will show the result in case 1 with these three improvements, whose details
were already explained in Sec.~\ref{sec:details}.

\subsection{Potential with additional noise reductions {in case 1}}
\begin{figure}[tbp]
  \hspace{-10mm}
  \begin{tabular}{cc}
  \begin{minipage}{0.5\hsize}
    \includegraphics[width=85mm,clip]{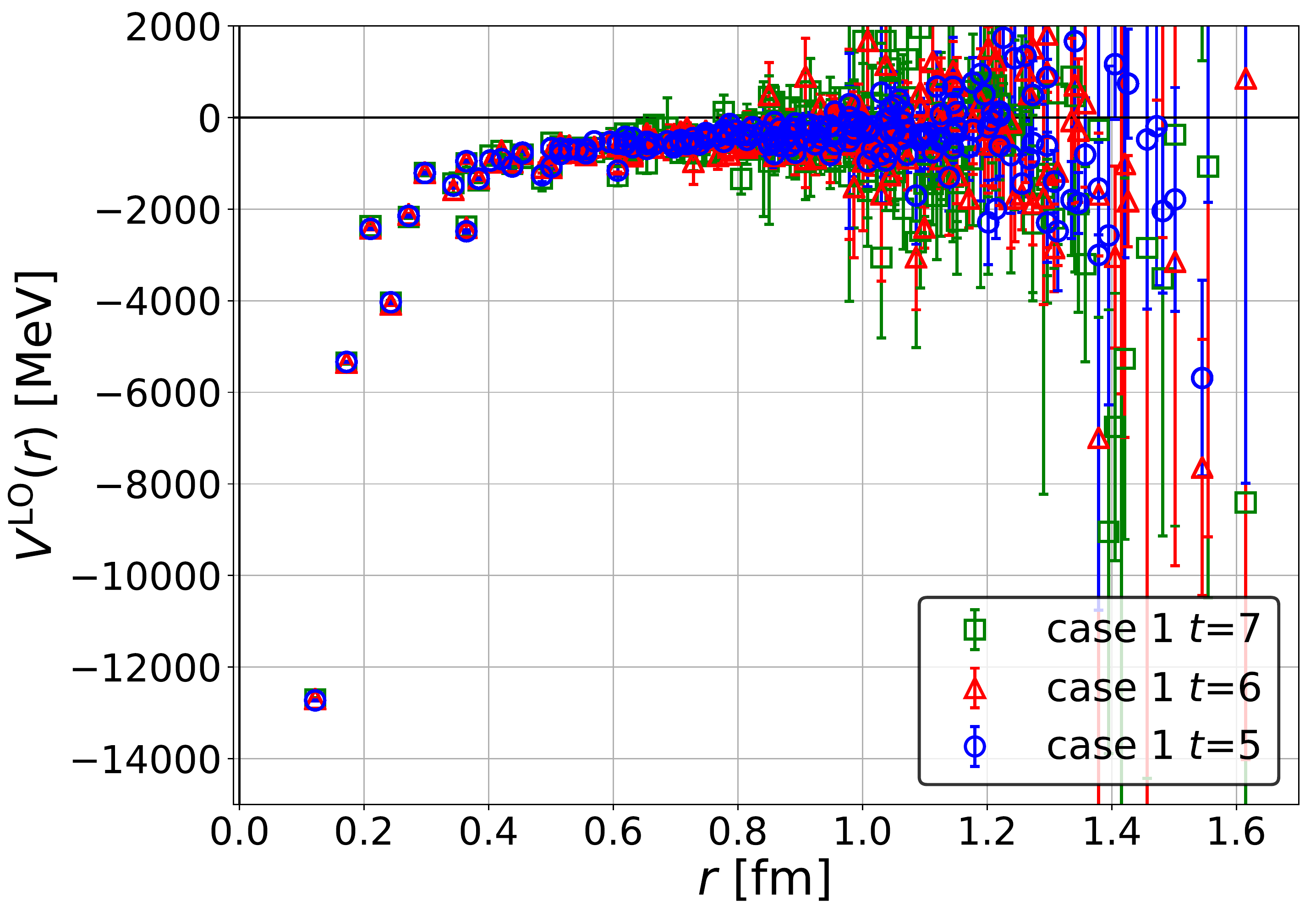}
  \end{minipage} &
  \begin{minipage}{0.5\hsize}
    \includegraphics[width=85mm,clip]{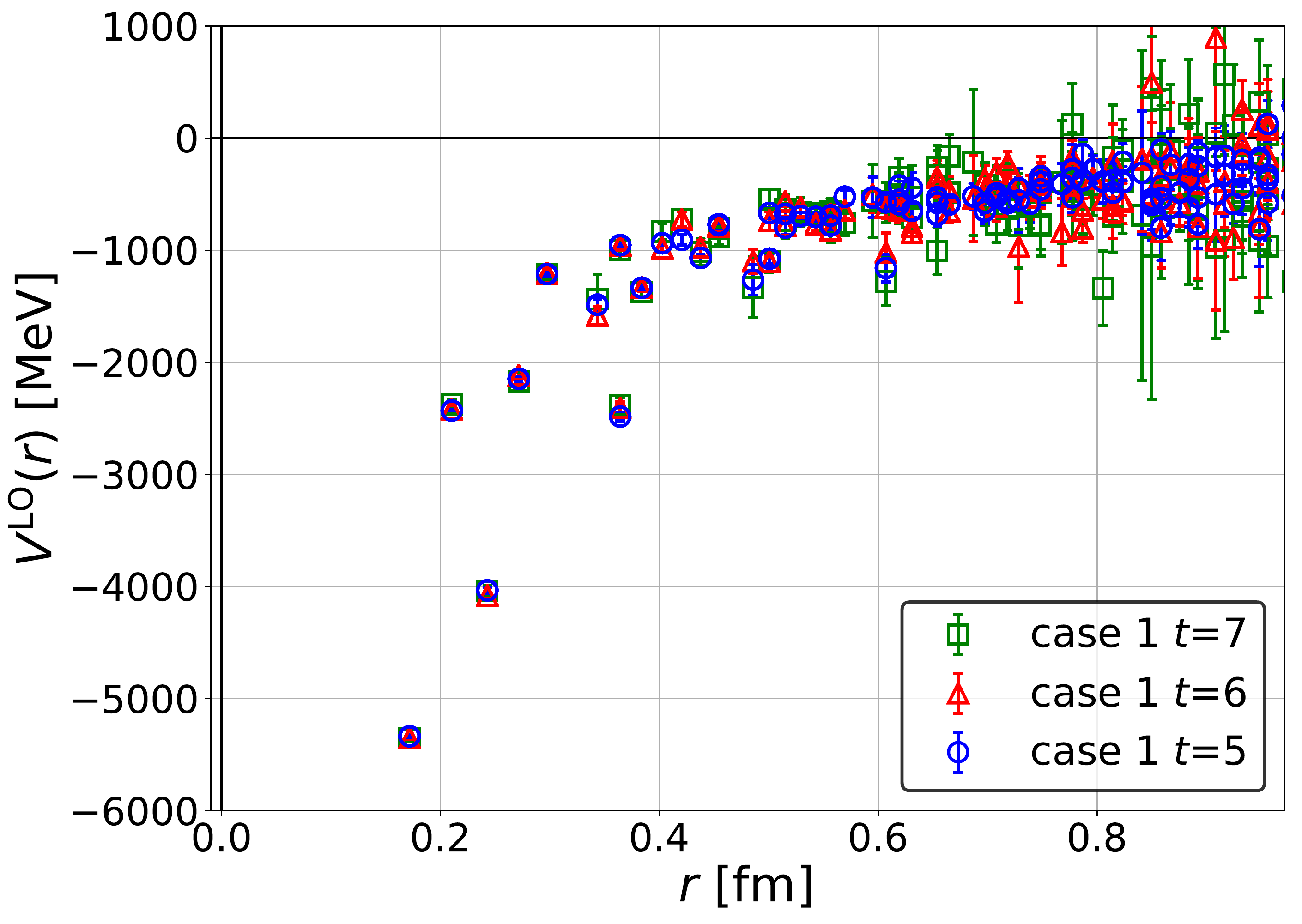}
  \end{minipage}
\end{tabular}
\caption{Time dependence of the potential in case 1. {(Left) Overall view. (Right) Enlarged view at $r \leq L/2$.} }
\label{fig:potentialtdep}
\end{figure}

The potential in case 1 at $t=6$ is shown in Fig.~\ref{fig:potential} (Right).
Since the smeared quark sources are employed in case 1, $t = 6$ is large enough
to suppress elastic contributions to the potential.
Thanks to additional noise reduction techniques mentioned in Sec.~\ref{sec:simple}, statistical fluctuations of the potential are drastically reduced.
The potential shows a strong attraction without repulsive core, which is consistent with existence of the deeply-bound $\rho$ meson in this system.
As shown in Fig.~\ref{fig:potentialtdep}, the potentials is almost independent of time
at $t=5,6,7$, as expected from
the effective energy shown in Fig.~\ref{fig:effmasses}.
Interestingly, we notice that statistical fluctuations of the potential increase as the distance $r$ increases.
We interpret this behavior qualitatively as follows.
Two-pion scattering states give dominant contributions to the long-distance part of the potential,
as the two-pion sink operator in the NBS wave function at large $r$ strongly couple to them.
The bound $\rho$ meson state, on the other hand, give large contributions to
the short distant part of the potential.
Since  the $\rho$-type operator we employ at the source hardly creates
such two-pion scattering states,
it is hard to determine the long-distance part of the potential precisely, and
thus statistical fluctuations become large.
We also observe that the short-distance part of the potential
has non-smooth behaviors, which probably
come from higher partial wave contaminations, for example, the $l=3$ partial wave in our case,
as similar behaviors have been sometimes observed for the HAL QCD potentials in previous studies
and the rotational breaking by the discretization artifact is expected to be enhanced at short-distance.

\begin{figure}[tbp]
  \centering
  \includegraphics[width=100mm,clip]{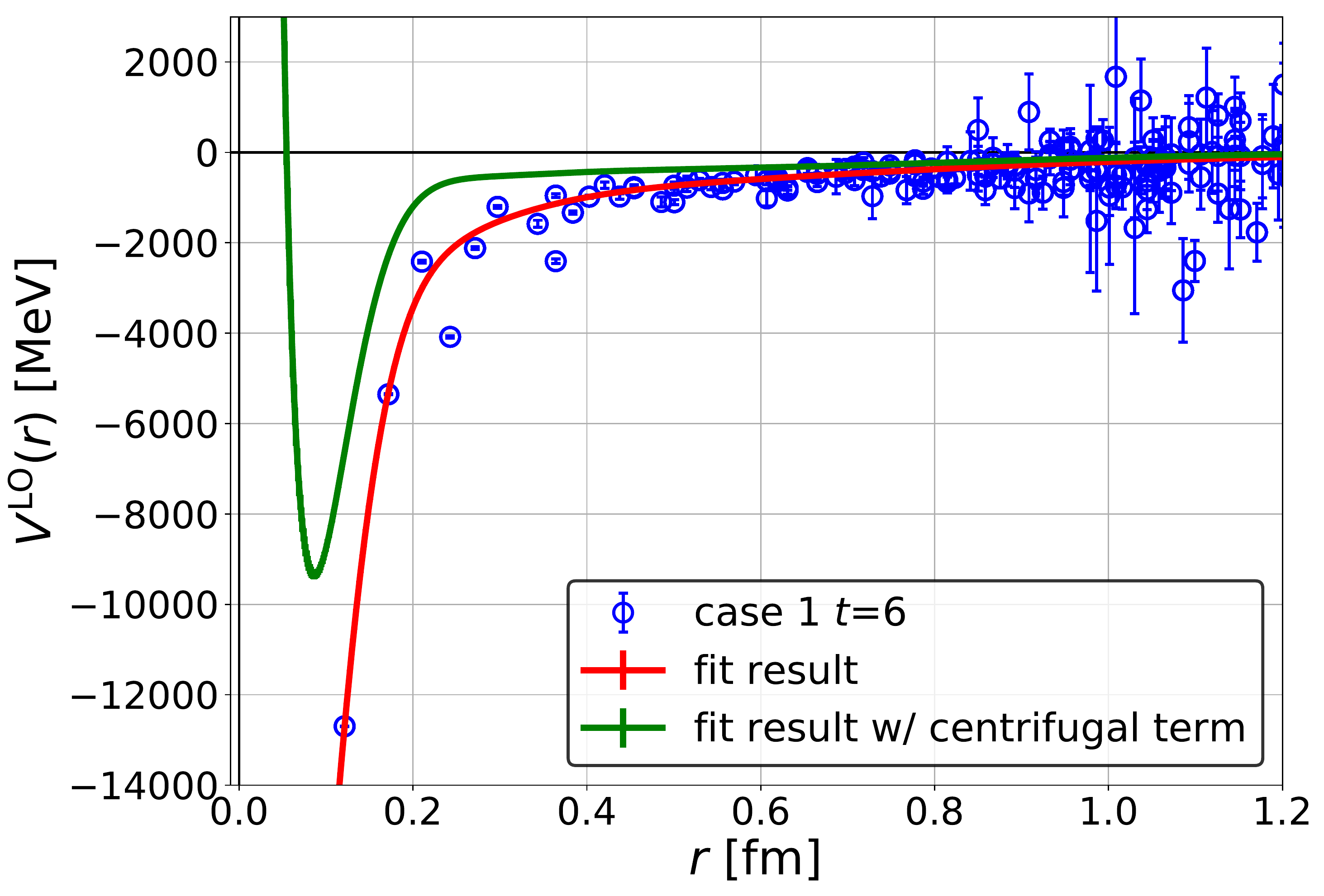}
  \caption{Fitting result at $t=6$. Blue points are the original data, and red line shows the fitting result. Green line is the fitting result curve with the centrifugal potential term with $l=1$, $V_c(r) = \frac{1}{m_{\pi}} \frac{1 \cdot 2}{r^2}$.}
  \label{fig:potentialfit}
\end{figure}

To calculate physical observables such as binding energies and scattering phase shifts, we fit the potential on discrete lattice points by a sum of {three} Gauss functions given by
\begin{equation} \label{eq:pot_fit}
  V(r) = a_0 e^{-(r/a_1)^2} + a_2 e^{-(r/a_3)^2} + a_4 e^{-(r/a_5)^2}.
\end{equation}
Several issues for the fit of the potential are in order here.
The first one is the finite volume effect.
As seen in Fig.~\ref{fig:potential} (Right),
the potential deviates from zero even at $r = La/2 = 0.9712$ fm
due to the finite volume effect of the periodic boundary condition.
We therefore partly include this finite volume effect into the fit as
\begin{equation}
  V({\bf r})_{\rm PBC} = V({\bf r}) + \sum_{{\bf n} \in \{(0,0,\pm1),(0,\pm1,0),(\pm1,0,0)\} } V({\bf r} + L {\bf n}).
  \label{eq:pot_PBC}
\end{equation}
The second issue is the non-smooth behavior of the potential at short distance, as mentioned before.
To make the fit stable, we have to exclude
two points of the potential at $r= 0.2428$ and $0.3642$ fm,
which largely deviate from other data points.
We expect that the exclusion of these points partly reduces the systematic uncertainty
associated with the contaminations from higher partial waves at short distances.
We leave a more detailed analysis for future investigations with {finer} lattices and
a new partial wave decomposition method~\cite{Miyamoto:2019jjc}.
Tab.~\ref{tab:fitparams} gives the result of the fit at $t=6$
and Fig.~\ref{fig:potentialfit} shows the original potential and the fitting result.
Note that the $\chi^2/$d.o.f.=7.59 is much larger than 1
even with the exclusion of two data points at $r= 0.2428$ and $0.3642$ fm in the fit,
since remaining data points at a short distance still have scattered central values with small statistical errors.

\begin{table}[tbp]
  \caption{Resultant fitting parameters and $\chi^2/d.o.f.$ at $t=6$. All values are in lattice unit.}
  \centering
  \begin{tabular}{cccccc|c}
    $a_0$ & $a_1$ & $a_2$ & $a_3$ & $a_4$ & $a_5$ & $\chi^2/d.o.f.$ \\ \hline \hline
    -1.7(0.2) & 2.0(0.1) & -0.64(0.04) & 6.5(0.2) & -21.0(0.1) & 0.886(0.008) & 7.59
  \end{tabular}
  \label{tab:fitparams}

\end{table}

\subsection{Physical observables}
{
Using the potential given by eq.~(\ref{eq:pot_fit}),
we calculate the ground state energy of the $I=1$ $\pi\pi$ system in the infinite volume.
We employ the Gaussian expansion method(GEM)~\cite{Hiyama:2003cu} to evaluate the ground state energy, which is given by
\begin{equation}
  E_{\rm bind} = 668 \pm 24_{\rm stat} \left( \begin{array}{c}
    + 69 \\ - 151
  \end{array}\right)_{\rm sys(time\ dep.)} {\rm MeV},
  \label{eq:B_energy}
\end{equation}
where the first error {denotes} the statistical error and the second error the systematic one estimated by the time dependence of the binding energy at $t=6 \pm 1$.
Comparing with the binding energy $E_{\rm bind} = |m_{\rho} - 2 m_{\pi}| \approx 515$ MeV from $m_{\pi}$ and $m_{\rho}$ (See Sec. 3),
{the results are consistent with each other within a large systematic error in eq.~(\ref{eq:B_energy}).}

{
  We also remark the systematic error associated with the fit of the potential.
  As mentioned in the previous subsection, some unreliable points have to be excluded in the fit,
  and data points at short distances are still scattered with small statistical errors,
  which leads to large $\chi^2/$d.o.f.
  In such a situation, 
  the fit at the short-range part as well as the resultant binding energy may have additional large uncertainty,
  since the latter is rather sensitive to the structure of the potential at short distances.
}
{
  While the corresponding systematic error is not fully quoted in eq.~(\ref{eq:B_energy}),
  part of such a systematics seems to be reflected in the systematic error estimated from the time dependence.
  In fact, we find that the time dependence of the results is substantial
  even though the potential is rather time independent as shown in Fig.~\ref{fig:potentialtdep}.
  This indicates that the large time dependence is mostly originated from the uncertainty in the fit of the potential.
}
To make systematic uncertainties {fully}
under control, we need to employ calculations at finer lattice spacings to {obtain} more data points at short distances or to find a better scheme for the NBS wave function to have smoother behaviors at short distances.
{Having remarked the above open issue,} 
we can {still} positively conclude that
it is possible to calculate reasonably precise potentials in the systems including quark creation$/$annihilation processes
by the combination of the hybrid method and the HAL QCD method.
}

We finally discuss a relation between $k^3 \cot \delta_1(k)$ and the bound state pole in detail,
as the {normality} check proposed in Ref.~\cite{Iritani:2017rlk}.
{
In the P-wave scattering, $k^3 \cot \delta_1(k)$ is related to the scattering S-matrix $S_1(k)$ as
\begin{equation} \label{eq:relation_k3cotd__smat}
  k^3 \cot \delta_1(k) = i k^3 \frac{S_1(k)+1}{S_1(k)-1}.
\end{equation}
Generally, the scattering S-matrix in P-wave near the bound state pole ($k \approx i\kappa_b$) behaves as\cite{Sitenko:scattering1991}
\begin{equation} \label{eq:smat_bound}
  S_1(k) \approx \frac{i \beta_b^2}{k-i\kappa_b},
\end{equation}
where $\kappa_b$ is an absolute value of $k$ of the pole and $\beta_b^2$ is positive real constant related to the normalization factor of the wave function of the bound state.
By using Eq.(\ref{eq:relation_k3cotd__smat}) and (\ref{eq:smat_bound}), the physical pole condition in P-wave becomes
\begin{equation} \label{eq:physical_pole_condition}
  \frac{d}{dk^2} \left. \left[ k^3 \cot \delta_1(k) - (-k^2 \sqrt{-k^2}) \right] \right|_{k^2 = -\kappa_b^2} = - \frac{\kappa_b^2}{\beta_b^2} < 0.
\end{equation}
{
In Fig.~\ref{fig:k3cotd_sqw}, we show typical behaviors of $k^3 \cot \delta_1(k)$ calculated by the square well potential in several cases. We can see how $k^3 \cot \delta_1(k)$ evolves when the attraction becomes stronger from Fig.~\ref{fig:k3cotd_sqw} (a) to Fig.~\ref{fig:k3cotd_sqw} (c).
As seen in Fig.~\ref{fig:k3cotd_sqw} (b),
the deeply-bound state appears as the intersection (blue solid star) between $-k^2 \sqrt{-k^2}$ (the bound state condition, black dashed line) and a branch of $k^3 \cot \delta_1(k)$ (red solid line) disconnected from a branch at the origin ($k^2=0$).}
Moreover, $k^3 \cot \delta_1(k)$ satisfies the physical pole condition, Eq.(\ref{eq:physical_pole_condition}) (See Fig.~\ref{fig:k3cotd_sqw} (b)(lower right)).

{
These two typical behaviors of $k^3 \cot \delta_1$ in the presence of one deeply-bound state
in the P-wave are indeed observed for our data obtained from the potential at $t=6$:
Fig~\ref{fig:k3cotd} (Left) shows that
an intersection between $-k^2 \sqrt{-k^2}$ (black dashed line) and
$k^3 \cot \delta_1 (k)$ (red solid line) in the branch disconnected from the origin
appears at $k^2/m_{\pi}^2 \approx -0.623$,
{corresponding to the GEM result, $E_{\rm bind} \approx 668$ MeV.}
{
  Shown in Fig~\ref{fig:k3cotd} (Right) is $(k^3 \cot \delta_1 (k) -(-k^2 \sqrt{-k^2}))/m_\pi^3$,
  and one can explicitly see how the physical pole condition is satisfied.
}
}
\begin{figure}[tbp]
  \hspace{-10mm}
  \begin{tabular}{cc}
  \begin{minipage}{0.5\hsize}
    \includegraphics[width=80mm,clip]{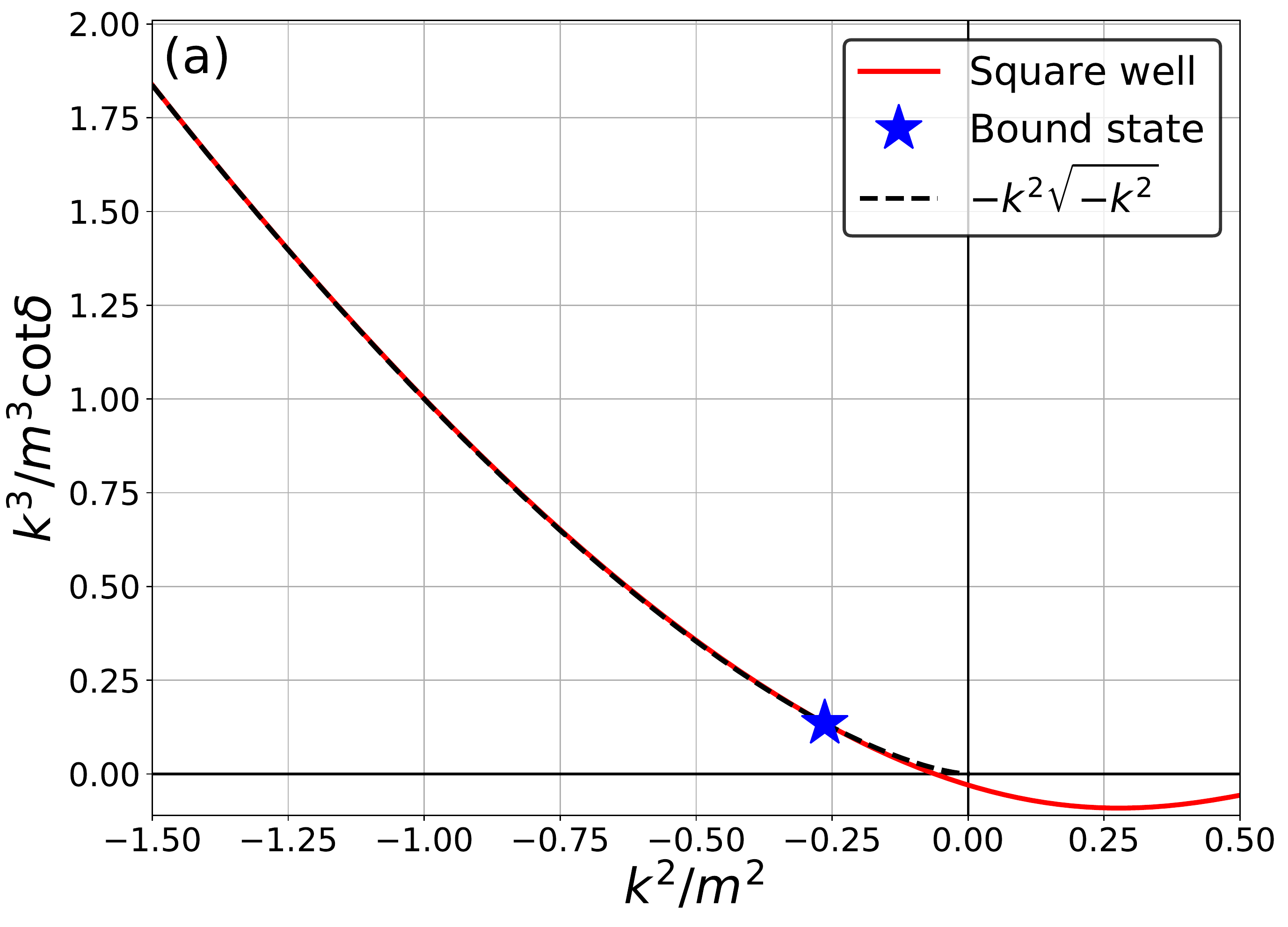}
  \end{minipage} &
  \begin{minipage}{0.5\hsize}
    \includegraphics[width=80mm,clip]{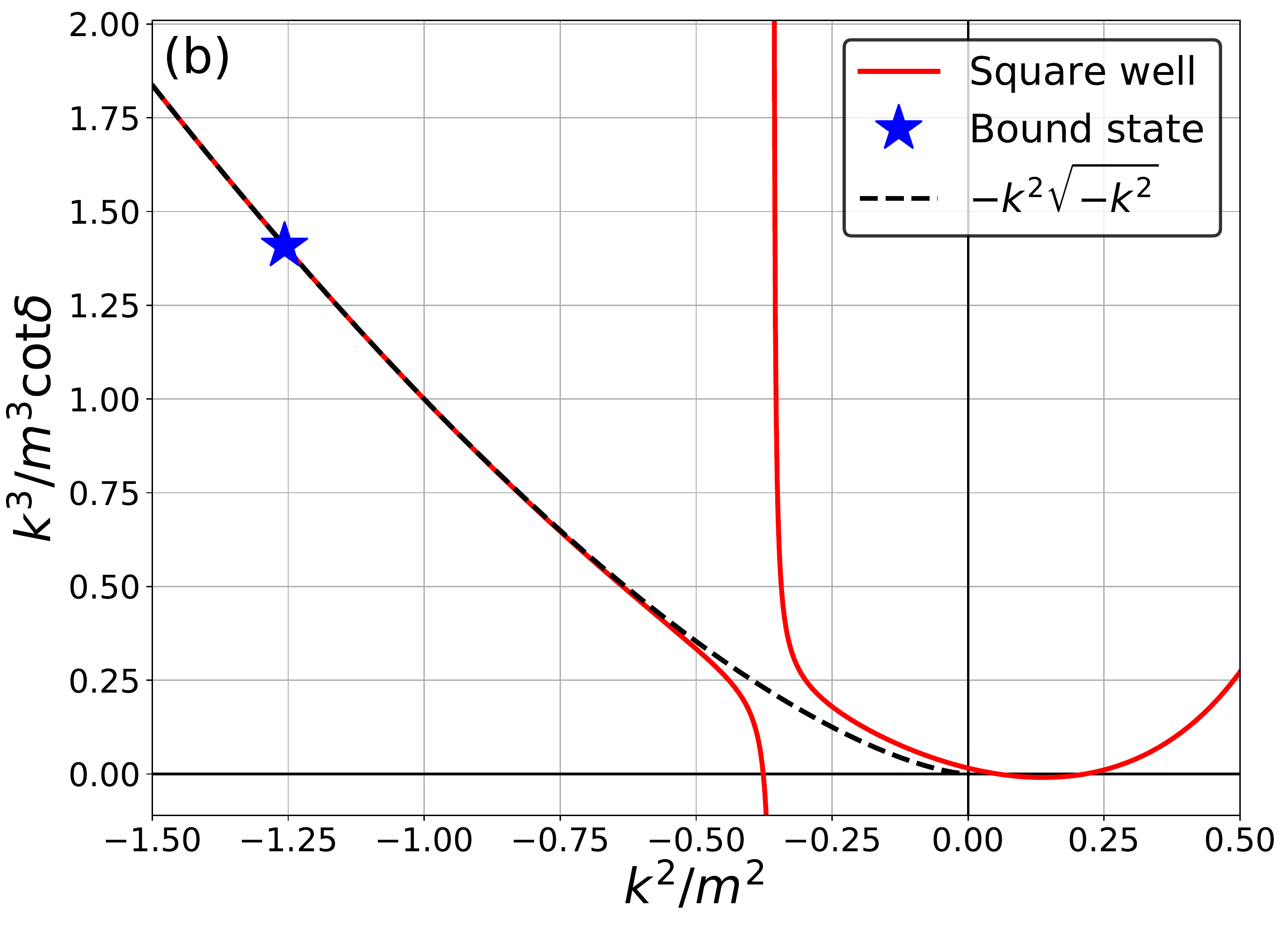}
  \end{minipage} \\
  \begin{minipage}{0.5\hsize}
    \includegraphics[width=80mm,clip]{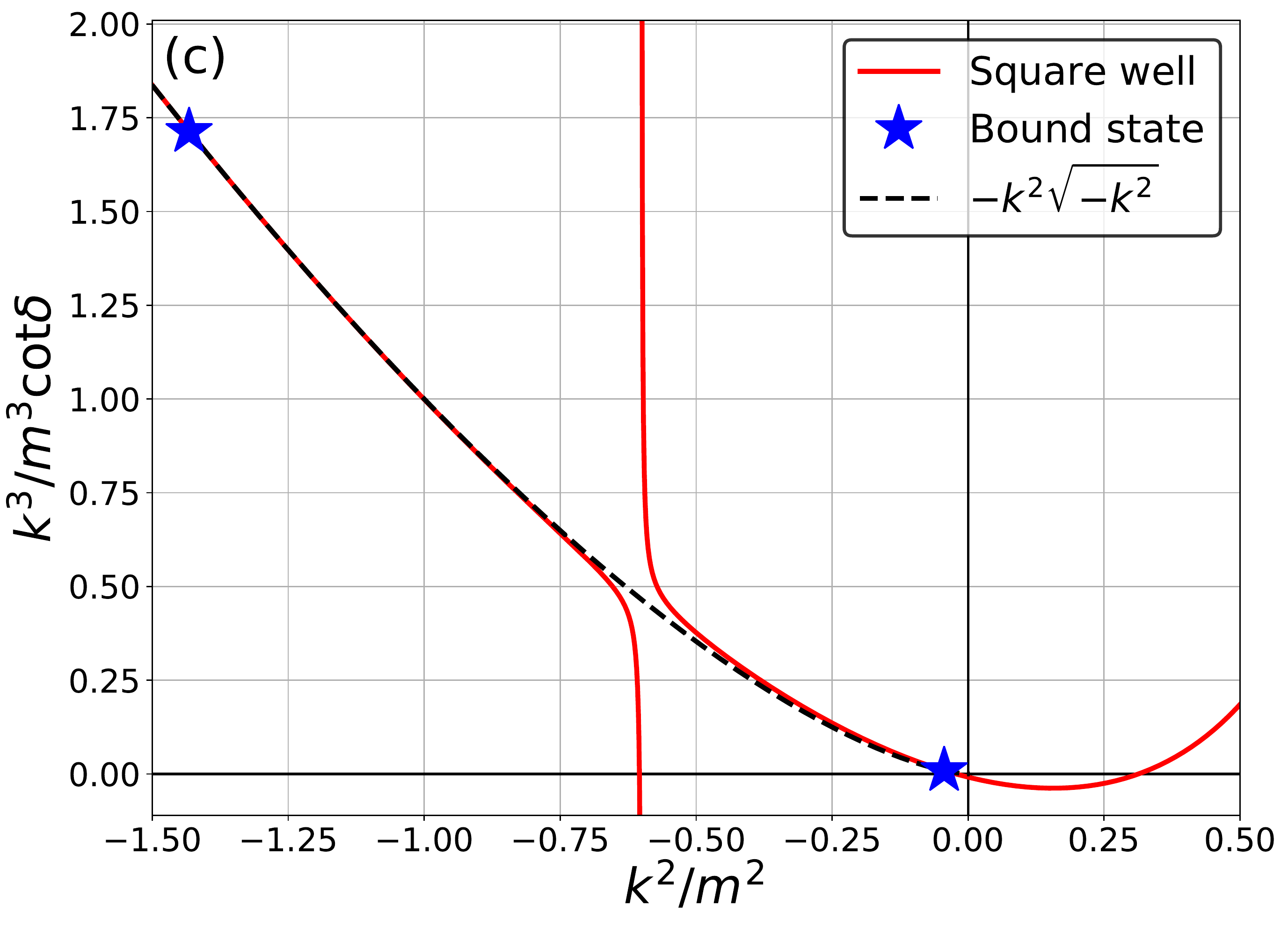}
  \end{minipage} &
  \hspace{-9mm}
  \begin{minipage}{0.5\hsize}
    \includegraphics[width=85mm,clip]{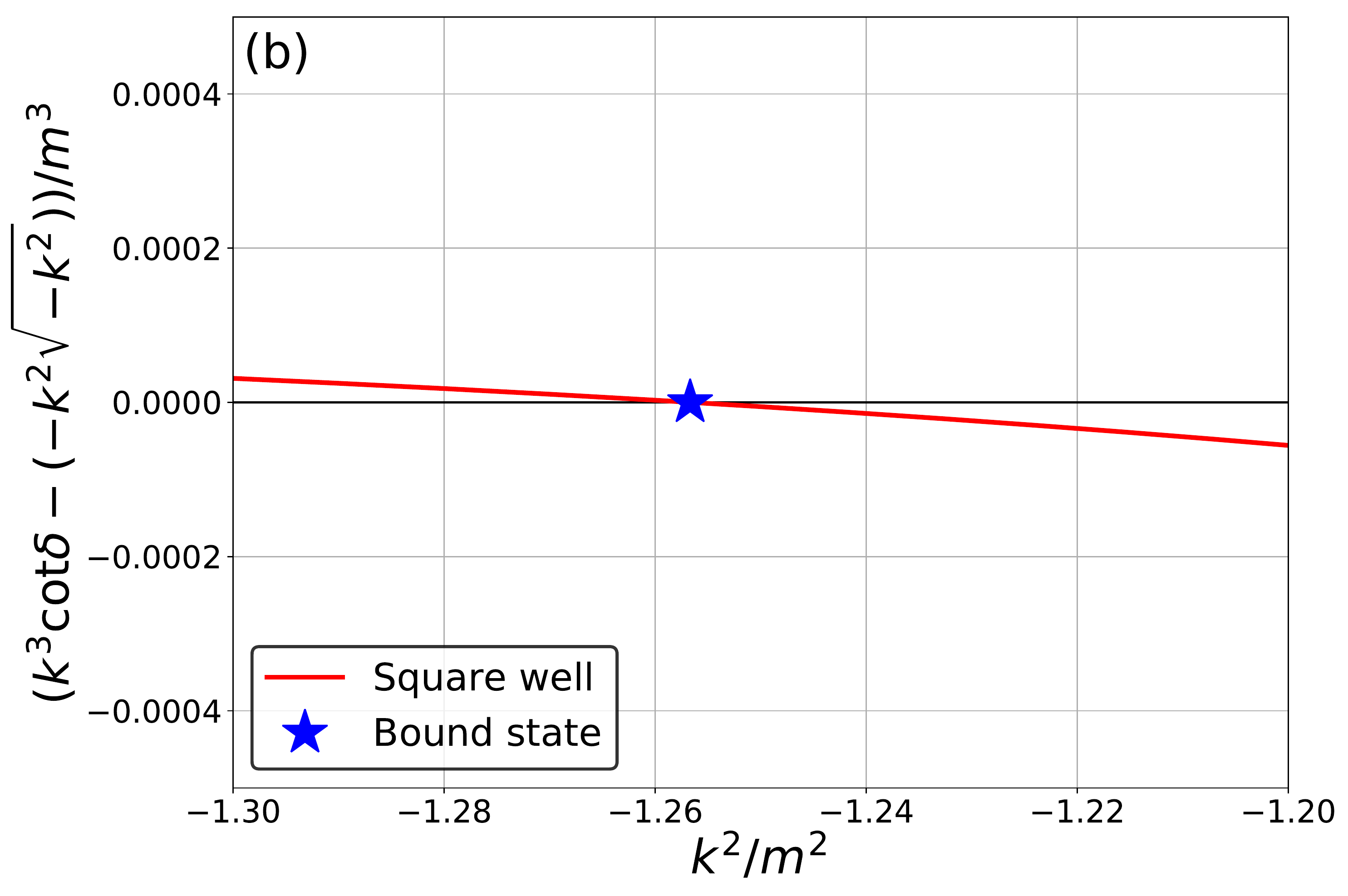}
  \end{minipage}
\end{tabular}
  \caption{A typical behavior of $k^3 \cot \delta_1$ in the P-wave calculated with the square well potential, together with the bound state condition (black dashed line) and pole of bound states (blue star). (a) shallowly-bound case (weak attraction). (b) deeply-bound case (strong attraction). The second branch can be seen on the right hand side. A difference between $k^3 \cot \delta_1(k)$ and $-k^2 \sqrt{-k^2}$ around the bound state pole is also shown on the lower right side. (c) doubly bound case (very strong attraction). The second branch reaches to the bound state condition and the second bound state emerges.}
  \label{fig:k3cotd_sqw}
\end{figure}

\begin{figure}[tbp]
  \hspace{-10mm}
  \begin{tabular}{cc}
  \begin{minipage}{0.5\hsize}
    \includegraphics[width=80mm,clip]{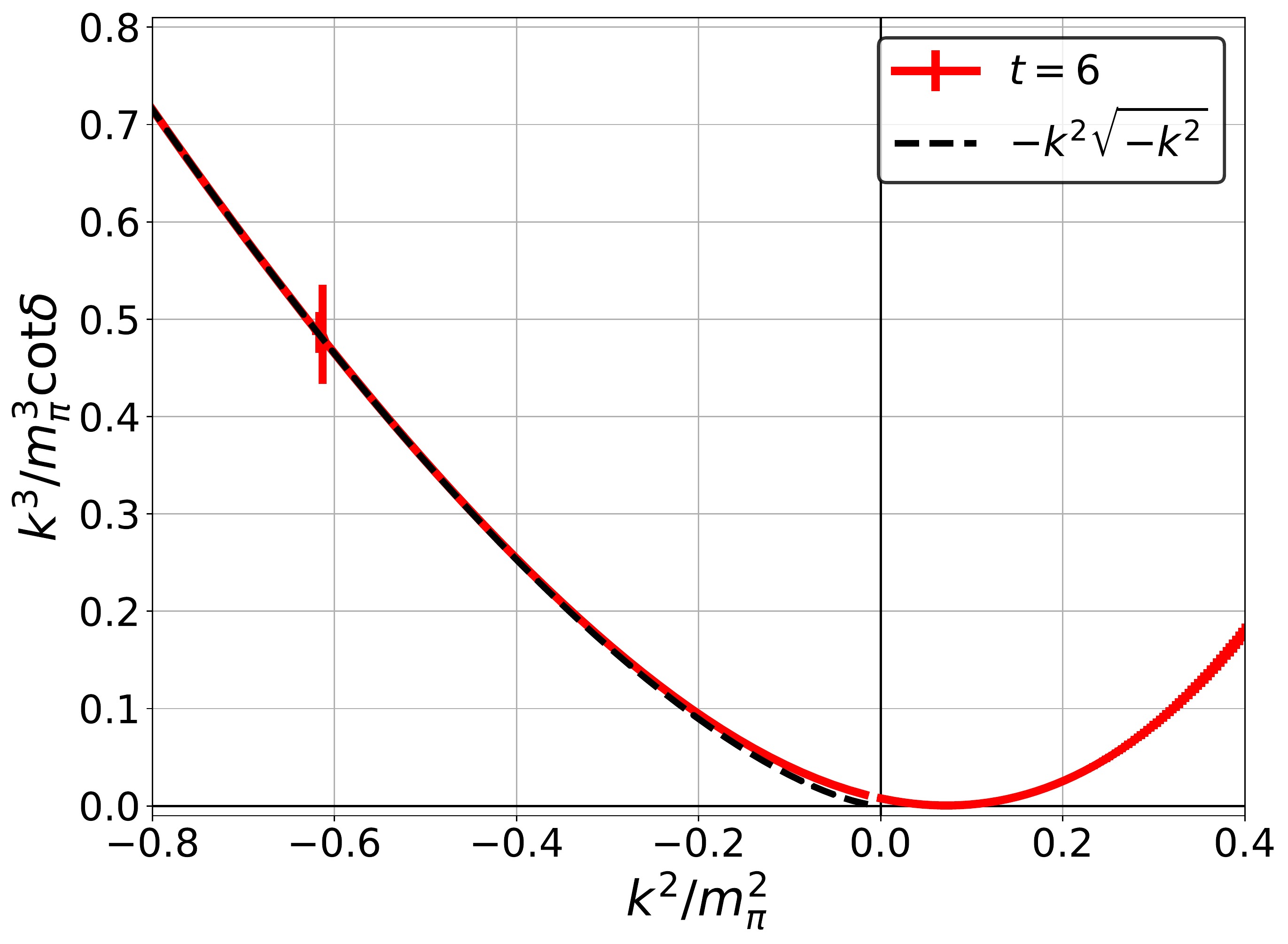}
  \end{minipage} &
  \hspace{-6mm}
  \begin{minipage}{0.5\hsize}
    \includegraphics[width=85mm,clip]{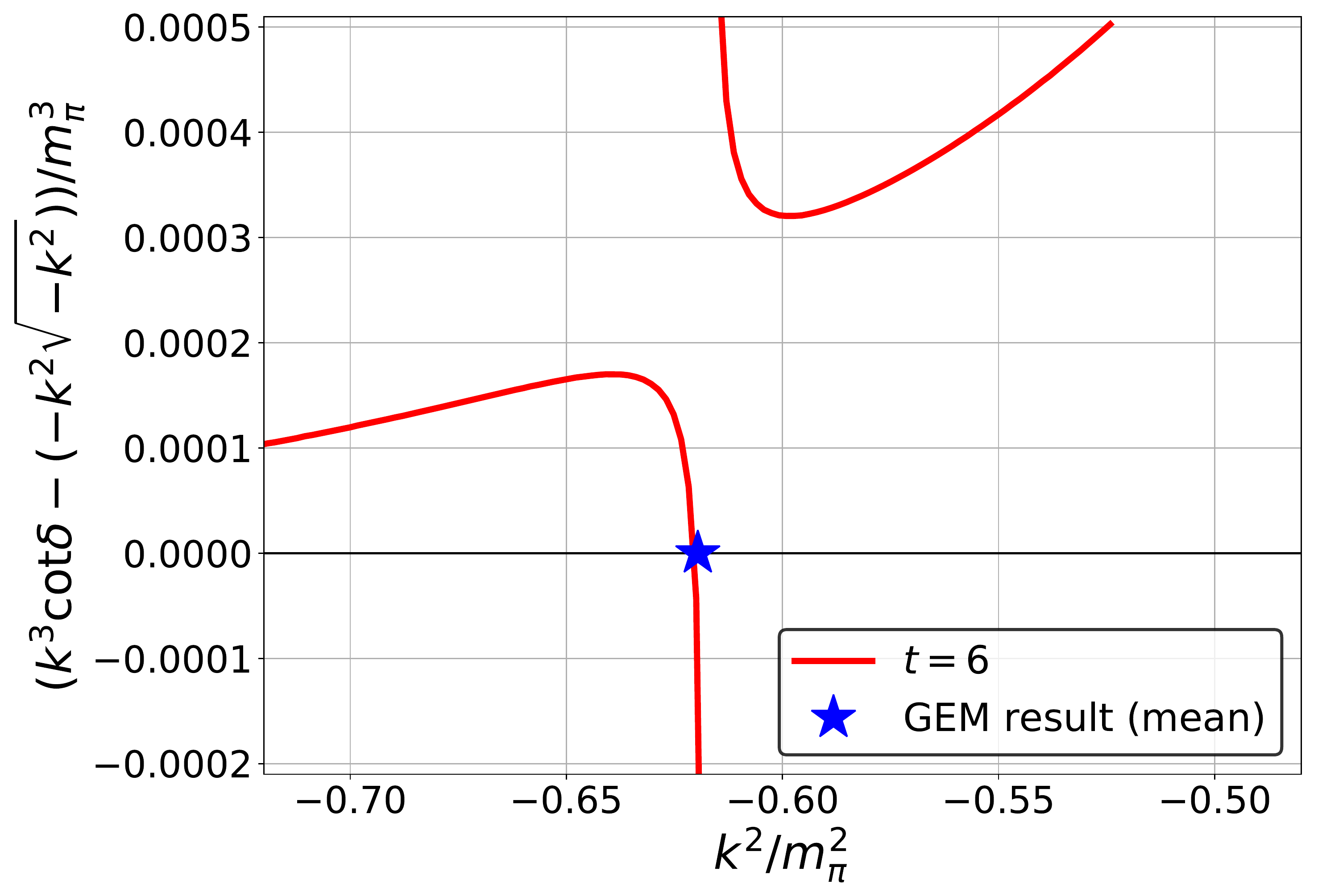}
  \end{minipage}
\end{tabular}
\caption{(Left) $k^3 \cot \delta_1(k)$ (red solid lines) calculated with the potential  at $t=6$,
together with the bound state condition (black dashed line). Note that red solid lines diverge to $\pm \infty$ around $k^2/m_{\pi}^2 \approx {-0.6}$. (Right) A difference between $k^3 \cot \delta_1(k)$ and $-k^2 \sqrt{-k^2}$ around the intersection, together with $k^2$ obtained by the Gaussian expansion method (blue solid star).
{Only the central values are used for the visibility.}
}
\label{fig:k3cotd}
\end{figure}


\section{Summary and outlook}
In this paper,
we calculate the HAL QCD potential of the $I=1$ $\pi \pi$ system at $(m_{\pi},m_{\rho}) \approx (870,1230)$ MeV, using the hybrid method for all-to-all propagators.
While statistical fluctuations {in the straightforward calculation}
are {found to be} extremely {large} due to the quark creation$/$annihilation process,
{we have successfully obtained the precise potential}
{by developing}
various noise reduction techniques such as space dilutions and the non-equal time scheme for the potential.
  We have calculated physical quantities such as the binding energy and phase shifts from the potential.
  It is observed that our potential reproduces the characteristic features of the deeply-bound $\rho$ meson,
  whose binding energy is consistent with that obtained from the temporal correlation within a large systematic error in the former.
  The large systematic error in the present calculations is caused by
  the uncertainty of the fit for the potential at short distances,
  whose origin is attributed to the contaminations from higher partial wave components.

Finally, we would like to comment on further improvements to our calculation in the future.
This and previous studies\cite{Akahoshi:2019klc} on the hybrid method reveals that one can obtain reasonably precise potentials as long as appropriate setups of calculations are introduced,
but on the other hand, it is also found that the numerical cost for noise reductions seems too large to perform such calculations on larger lattice volumes.
Therefore, we have to investigate possibilities to achieve both small noise contamination and small numerical costs.
Fortunately, we find that the combination of some other techniques such as the all-mode-averaging\cite{Shintani:2014vja}, the one-end trick and sequential propagators\cite{Abdel-Rehim:2017dok} is promising to achieve above requirements.
As a first step toward this direction,
we are now working on the $\rho$ resonance at $m_{\pi} \approx 410$ MeV with new improved methods, and results will be reported in near future.

\section{Acknowledgement}
The authors thank members of the HAL QCD Collaboration for fruitful discussions. We thank the JLQCD and CP-PACS Collaborations~\cite{Ishikawa:2007nn} and ILDG/JLDG~\cite{Amagasa:2015zwb} for providing their configurations. All of the simulations are performed on the HOKUSAI Big-Waterfall in RIKEN. The framework of our numerical code is based on Bridge++ codeset~\cite{Ueda:2014rya}.
This work is supported in part by the Grant-in-Aid of the Japanese Ministry of Education, Sciences and Technology, Sports and Culture (MEXT) for Scientific Research (Nos. JP16H03978, JP18H05236, {JP18H05407, JP19K03879}),
by a priority issue (Elucidation of the fundamental laws and evolution of the universe) to be tackled by using Post ``K" Computer, and by Joint Institute for Computational Fundamental Science (JICFuS).


\end{document}